\documentclass[12pt,english]{extarticle}
\usepackage{mathptmx}
\usepackage{helvet}

\usepackage[T1]{fontenc}
\usepackage[latin9]{inputenc}
\usepackage{geometry}
\usepackage{color}
\usepackage{enumitem}
\usepackage{amsmath}
\usepackage{amsthm}
\usepackage{amssymb}
\usepackage{setspace}
\makeatletter
\IfFileExists{candleshape.def}{%
 \input{candleshape.def}}{}
\IfFileExists{dropshape.def}{%
 \input{dropshape.def}}{}
\IfFileExists{TeXshape.def}{%
 \input{TeXshape.def}}{}
\IfFileExists{triangleshapes.def}{%
 \input{triangleshapes.def}}{}

\newlength{\lyxlabelwidth}      % auxiliary length 
\@ifundefined{lettrine}{\usepackage{lettrine}}{}
\numberwithin{equation}{section}
\numberwithin{figure}{section}
\numberwithin{table}{section}
	\newenvironment{elabeling}[2][]%
	{\settowidth{\lyxlabelwidth}{#2}
		\begin{description}[font=\normalfont,style=sameline,
			leftmargin=\lyxlabelwidth,#1]}
	{\end{description}}
\theoremstyle{definition}
\ifx\thechapter\undefined
  \newtheorem{example}{\protect\examplename}
\else
  \newtheorem{example}{\protect\examplename}[chapter]
\fi

\makeatother

\usepackage{babel}
\providecommand{\examplename}{Example}

\begin{document}
\title{Inconsistency of Score-Elevated Reserve Policy for Indian Affirmative
Action\thanks{First version: March, 2023. Some of the  results in this paper was
part of an earlier working paper titled ``\emph{Affirmative Action
in India via Backward and Forward Transfers}.'' We are grateful to\emph{
}Battal Do\u{g}an, Isa Hafal\i r, and Bumin Yenmez for their feedback.
We acknowledge that we used ChatGPT and Grammarly to improve the readability
and language of the introduction section. }}
\author{Orhan Aygün\thanks{orhan.aygun@boun.edu.tr; University of Minnesota, Applied Economics
Department, 231 Ruttan Hall, 1994 Buford Ave, St. Paul, MN 55108,
USA; and Bo\u{g}aziçi University, Department of Economics, Natuk Birkan
Building, Bebek, Istanbul 34342, Turkey.} $\quad$and $\quad$Bertan Turhan\thanks{bertan@iastate.edu; Iowa State University, Department of Economics,
Heady Hall, 518 Farm House Lane, Ames, IA 50011, USA.}}
\date{December, 2023}
\maketitle
\begin{abstract}
India has enacted an intricate affirmative action program through
a reservation system since the 1950s. Notably, in 2008, a historic
judgment by the Supreme Court of India (SCI) in the case of Ashoka
Kumar Thakur vs. Union of India mandated a 27 percent reservation
to the Other Backward Classes (OBC). The SCI's ruling suggested implementing
the OBC reservation as a soft reserve without defining a procedural
framework. The SCI recommended a maximum of 10 points difference between
the cutoff scores of the open-category and OBC positions. We show
that this directive conflicts with India's fundamental Supreme Court
mandates on reservation policy. Moreover, we show that the score-elevated
reserve policy proposed by Sönmez and Ünver (2022) is inconsistent
with this directive.

\vfill{}
\end{abstract}
\begin{elabeling}{00.00.0000}
\item [{$\mathbf{Keywords:}$}] Market design, reserve systems, de-reservation,
hard reserves, soft reserves, score-elevated reserve policy, India.
\end{elabeling}
$\mathbf{JEL\;Codes}$\emph{: }C78, D47, D63.\pagebreak{}

\section{Introduction}

India has enforced the most intricate affirmative action program in
the world to allocate government jobs and public university seats
since 1950s via a highly regulated reservation system that consists
of \emph{vertical} and \emph{horizontal} reservations. These terms
were first used by the Supreme Court of India (SCI) in a historic
judgment in \emph{Indra Shawney vs. Union of India} (1992), henceforth
Indra Shawney (1992).\footnote{The case is available at https://indiankanoon.org/doc/1363234/.}
After the verdict in Indra Shawney (1992), the interplay between vertical
and horizontal reservations have been refined through consecutive
SCI decisions. 

By established legal framework, reservations for \emph{Scheduled Castes}
(SC), \emph{Scheduled Tribes} (ST), \emph{Other Backward Classes}
(OBC), and \emph{Economically Weaker Sections} (EWS) are categorized
as \textbf{vertical reservations}. Under this schema, 15\%, 7.5\%,
27\%, and 10\% of available institutional positions are set aside
for these respective social strata.\footnote{In 2019, the Indian legislative body instituted a 10\% vertical reservation
for a subset of the \emph{General Category} (GC), specifically the
Economically Weaker Section, defined by an annual income threshold
below Rs. 8 lakhs.} Individuals not encompassed by these vertical categories are classified
as members of the \emph{General Category} (GC). The residual 40.5\%
of available positions fall under the \emph{open category}. The governing
statutes stipulate that any open-category positions secured by individuals
belonging to SC, ST, OBC, or EWS are not to be deducted from their
respective vertical reservations. Hence, vertical reservations operate
under an ``\textbf{over-and-above}\textquotedbl{} legal principle.
Failure to declare membership in SC, ST, OBC, or EWS categories relegates
candidates to the GC by default. Importantly, the disclosure of vertical
category affiliation is discretionary. 

In governmental employment and admissions to publicly funded academic
institutions, reservations for SC and ST have been introduced and
implemented as \textbf{hard} reserves, meaning unfilled positions
earmarked for these groups are non-transferable to other categories.
While the reservations for OBC have also been treated as hard reserves
in the context of governmental job allocations, they were suggested
and have been implemented as \textbf{soft} reserves in admissions
to public universities, according to the landmark SCI decision in
\emph{Ashoka Kumar Thakur vs. Union of India} (2008).\footnote{The judgment is available at https://indiankanoon.org/doc/1219385/.}
The judgment in Ashoka Kumar Thakur (2008) upheld the validity of
27\% reservation for OBCs in educational institutions and government
jobs. However, it did not proffer explicit procedural details for
reverting unfilled OBC positions to others. 

The absence of judicial guidance has caused considerable ambiguity
concerning the eligible beneficiaries of the de-reserved OBC positions---whether
they should be exclusively limited to GC candidates or be open to
all, including those in reserved categories. The lack of clear rules
has led to various inconsistent practices in school admissions across
India (Aygün and Turhan 2022). The ruling in Ashoka Kumar Thakur (2008)
states 
\begin{quote}
\emph{Under such a scheme, whenever the non-creamy layer OBCs fail
to fill the 27\% reservation, the remaining seats would revert to
general category students.}
\end{quote}
In the same judgment, the SCI examined several facets of reservation
policies, especially the concept of \textbf{cut-off scores} for different
categories. The court explicitly stated that there should be a \emph{reasonable
difference} in the cut-off scores between the OBC and open categories
as follows: 
\begin{quote}
\emph{It is reasonable to balance reservation with other societal
interests. To maintain standards of excellence, cut-off marks for
OBCs should be set not more than 10 marks out of 100 below that of
the general category.}
\end{quote}
In the first version of their paper\footnote{The paper is available at https://arxiv.org/abs/2210.10166v1.}---referring
to the SCI directive in the judgment of Ashoka Kumar Thakur (2008)
given above---Sönmez and Ünver (2022) interpret the OBC de-reservation
policy such that it gives an additional 10 points to the members of
the OBC category and keeps everyone eligible for the OBC category
positions.\footnote{Based on this directive, in the \textbf{first version} of their paper,
\textcolor{black}{Sönmez and Ünver (2022)} criticized the \emph{Backward
Transfers} (BT) choice rule introduced by Aygün and Turhan (2022)
as follows: 
\begin{quote}
\emph{The authors propose to accommodate the above-given directives
of the Supreme Court by (i) adopting a soft VR-protection policy for
OBC and (ii) processing the OBC category prior to all other VR-protected
categories. Since replacing the regular VR protection policy with
the soft VR protection policy for OBC affects the outcome only when
there isn't a sufficient number of applicants (regardless of the merit
scores of existing OBC applicants), the choice rule proposed by Aygün
and Turhan (2022) fails to accommodate the above-given directive of
the Supreme Court. Moreover, as we present in Theorem 2, the order
of precedence for the OBC proposed by Aygün and Turhan (2022) also
does not serve the stated objective of the Supreme Court in maintaining
standards of excellence. As a result, we advocate for a completely
different choice rule.}
\end{quote}
\textcolor{black}{Baswana et al. (2019) reports that they suggested
to modify the priority ordering for the OBC category by ranking all
non-OBC applicants below the OBC applicants according to the merit
scores. The authorities did }\textbf{\textcolor{black}{not}}\textcolor{black}{{}
accept this proposal because they feared this approach would cause
a lower cutoff score for the open category than the OBC category.
The design team then offered a process that transfers vacant OBC slots
to the open category in a backward manner by re-running the DA algorithm
on all applicants. Aygün and Turhan (2022) designed the BT choice
rule to better implement the policy. Sönmez and Ünver (2022) fail
to mention that the BT choice rule's underlying design objective is
to obtain a higher cutoff score for the open category than for reserved
categories. }} \textcolor{black}{Sönmez and Ünver (2022) }reads 
\begin{quote}
\emph{For the case of seat allocation at elite educational institutions,
this judgment provided a directive that corresponds to giving a 10
points of advantage to beneficiaries of the OBC category for allocation
of VR-protected positions for this category, rather than exclusively
reserving these positions for them. Motivated with this directive,
we present a general model of reservation allows both for overlapping
VR protections, and also for non-regular VR policies. Moreover, in
relation to the directive of Ashoka Kumar Thakur (2008), in Theorem
2 we show that the best policy for maintaining standards of excellence
can be achieved by processing OBC-category positions after all other
VR-protected categories.}
\end{quote}
In this paper, we show that the recommendation provided in Ashoka
Kumar Thakur (2008)---imposing certain maximum difference between
the open category and OBC cutoff scores---conflicts with India's
fundamental mandates on reservation policy introduced in Indra Shawney
(1992). Moreover, we prove that \textbf{the score-elevated reserve
policy} proposed by Sönmez and Ünver (2022) is inconsistent with this
directive. 

\section{The Preliminaries}

We adopt the notation of the first version of Sönmez and Ünver (2022)
for consistency. 

There is a finite set $\mathcal{I}$ individuals. Each individual
$i\in\mathcal{I}$ has a unit demand. The function $\sigma:\mathcal{I}\rightarrow\mathbb{R}_{+}$
denotes individuals' test scores. We denote by $\sigma=\left(\sigma_{i}\right)_{i\in\mathcal{I}}$
the vector of individuals' test scores. 

Let $\mathcal{R}=\left\{ SC,ST,OBC,EWS\right\} $ denote the vertical
reserve categories. An individual who is not a member of any vertical
reserve category is a member of general category $g\notin\mathcal{R}$.
The correspondence $\rho:\mathcal{I}\rightrightarrows\mathcal{R}\cup\left\{ g\right\} $
denotes the vertical category membership of individuals. The correspondence
$\rho$ is such that $g\in\rho\left(i\right)$ if and only if $\rho\left(i\right)=g$,
which means that individual $i$ is not a member of any reserve category.
The set $\mathcal{I}^{r}\left(\rho\right)=\left\{ i\in\mathcal{I}\mid r\in\rho\left(i\right)\right\} $
denotes the set of vertical category-r members. The set $\mathcal{I}^{g}=\mathcal{I}\setminus\bigcup_{r\in\mathcal{R}}\mathcal{I}^{r}$
denotes the set of individuals in the general category. 

There is a single institution with $q$ positions. $q^{r}$ positions
are set aside for vertical category $r$ members. We assume that $\underset{r\in\mathcal{R}}{\sum}q^{r}\leq q$.
The remaining $q^{o}=q-\underset{r\in\mathcal{R}}{\sum}q^{r}$ positions
are \textbf{open category }positions. All individuals are eligible
for open category positions. 

Individuals test scores induce the \textbf{baseline priority order
}$\pi^{o}$ over the set $\mathcal{I}\cup\left\{ \emptyset\right\} $.
The baseline priority order $\pi^{o}$ is such that 
\[
i\pi^{o}j\pi^{o}\emptyset\quad\Longleftrightarrow\quad\sigma\left(i\right)>\sigma\left(j\right)
\]
 for any two individuals $i,j\in\mathcal{I}$ with $i\neq j$. 

\section{Case Against the Score-Elevated Policy in Indian Affirmative Action}

Sönmez and Ünver (2022) formulated and advocated the following \emph{score-elevated
vertical reservation policy} as a better alternative in Section 2.1.1
of the first version of their paper: 
\begin{quotation}
\emph{``Given a positive number $k\in\mathbb{R}_{+}$, (that is interpreted
as the amount of a boost for merit scores of the members of category
c), our third VR protection policy $\overset{\frown}{\pi}^{c}$ is
defined by the following properties:}

\emph{(1) For any pair of distinct individuals, $i,j\in\mathcal{I}^{c}\left(\rho\right)$,
\[
i\overset{\frown}{\pi}^{c}j\overset{\frown}{\pi}^{c}\emptyset\;\Longleftrightarrow\;\sigma_{i}>\sigma_{j}.
\]
}

\emph{(2) For any pair of distinct individuals $i,j\in\mathcal{I}\setminus\mathcal{I}^{c}\left(\rho\right)$,
\[
i\overset{\frown}{\pi}^{c}j\overset{\frown}{\pi}^{c}\emptyset\;\Longleftrightarrow\;\sigma_{i}>\sigma_{j}.
\]
}

\emph{(3) For any pair of distinct individuals $i\in\mathcal{I}\left(\rho\right)$
and $j\in\mathcal{I}\setminus\mathcal{I}^{c}\left(\rho\right)$, 
\[
i\overset{\frown}{\pi}^{c}j\;\Longleftrightarrow\;\sigma_{i}+k>\sigma_{j}."
\]
}
\end{quotation}
We show with a simple example that the policy they advocated is\emph{
inconsistent }with the directive they refer to. 
\begin{example}
Consider an institution with four positions. The institution reserves
one positions for each of SC, ST, and OBC categories. The remaining
position is unreserved. There are five individuals with the following
scores and category membership: 

\[
\begin{array}{ccc}
Individual & Category & Score\\
i_{1} & g & 100\\
i_{2} & SC & 99\\
i_{3} & ST & 98\\
i_{4} & g & 98\\
i_{5} & OBC & 89
\end{array}
\]

According to Sönmez and Ünver's (2022) score-elevated reservation
policy with $k=10$, the chosen set of individuals is $\left\{ i_{1},i_{2},i_{3},i_{5}\right\} $
because 10 points score-elevation leads to score of 99 for the OBC
individual $i_{5}$, which exceeds the score of the individual $i_{4}$
with $\rho\left(i_{4}\right)=g$. According to the chosen set, the
cutoff score for the open category is 100 and the cutoff score for
the OBC is 89. The difference between them is 11, which exceeds the
maximum allowable difference $k=10$. 
\end{example}
This example shows that the score-elevated reserve policy fails to
implement the directive the authors referred to.  The complication arises from accepting the OBC individual
with an elevated score 99, which exceeds the score of the general
category individual with a score of 98. 

\section{The Impropriety of the Ashoka Kumar Thakur (2008) Directive }

If one insists to keep the differences in cutoff scores of open category
and OBC at no more than 10, it may create complementarities between
different individuals. Therefore, the directive itself is problematic.
We show it with the following example.
\begin{example}
Consider an institution with 5 positions. One position is reserved
for SC, one for ST, and two for OBC. The remaining position is open
category. There are six individuals with the following scores and
category membership: 

\[
\begin{array}{ccc}
Individual & Category & Score\\
i_{1} & g & 100\\
i_{2} & SC & 99\\
i_{3} & ST & 98\\
i_{4} & OBC & 91\\
i_{5} & OBC & 90\\
i_{6} & g & 98
\end{array}
\]

Assuming within category fairness, non-wastefulness, and one of the
VR protection policy given in Sönmez and Ünver (2022), to keep the
difference between the open category and OBC cutoffs at 10, the chosen
set must be $\left\{ i_{1},i_{2},i_{3},i_{4},i_{5}\right\} $. The
open category cutoff is 100, and the OBC cutoff is 90. 

Now, suppose a new general category individual $i_{7}$ with a score
of 102 arrives. The chosen set from the set of individuals $\left\{ i_{1},i_{2},i_{3},i_{4},i_{5},i_{6},i_{7}\right\} $
is found as follows: $i_{7}$ is chosen for the open category position.
$i_{2}$ and $i_{3}$ are chosen for the SC and ST positions, respectively.
Since the  cutoff score is now 102, the OBC cutoff score cannot be
lower than 92. Therefore, for the remaining two OBC positions, $i_{1}$
and $i_{6}$ are chosen. Note that $i_{6}$ was rejected before the
arrival of $i_{7}$, and now she is chosen. This violates the substitutes
property. Inclusion of $i_{7}$ helps $i_{6}$ to be chosen. 
\end{example}
Therefore, not only the score-elevated reserve policy of Sönmez and
Ünver (2022) may fail to implement the directive they are referring
to, but also the directive itself along with the score-elevation interpretation
of it creates complementarities among individuals. Hence, it violates
the crucial substitutes condition.

\end{document}